\documentclass[12pt]{iopart}
  \usepackage[dvips]{graphicx}
  \usepackage{harvard}
\usepackage{color}

\newcommand{\leg}[1]{\textbf{#1}}

\begin{document}
\graphicspath{{./FIGURES/}}

\paper[Wind statistics modeling and wind power evaluation]{Is the Weibull distribution really suited for wind statistics modeling and wind power evaluation?}

\author{P. Drobinski$^1$ and C. Coulais$^2$}
\address{$^1$ Laboratoire de M\'{e}t\'{e}orologie Dynamique - Institut Pierre
Simon Laplace, CNRS and Ecole Polytechnique, Palaiseau, France}
\address{$^2$ CEA Saclay/DSM/IRAMIS/SPEC/SPHYNX, 91191 CEDEX Gif-sur-Yvette, France}
\ead{philippe.drobinski@lmd.polytechnique.fr}

\begin{abstract}
Wind speed statistics is generally modeled using the Weibull distribution. This distribution is convenient since it fully characterizes analytically with only two parameters (the shape and scale parameters) the shape of distribution and the different moments of the wind speed (mean, standard deviation, skewness and kurtosis). This distribution is broadly used in the wind energy sector to produce maps of wind energy potential. However, the Weibull distribution is based on empirical rather than physical justification and might display strong limitations for its applications. The philosophy of this article is based on the modeling of the wind components instead of the wind speed itself. This provides more physical insights on the validity domain of the Weibull distribution as a possible relevant model for wind statistics and the quantification of the error made by using such a distribution. We thereby propose alternative expressions of more suited wind speed distribution.
\end{abstract}
\submitto{Environmental Research Letters}


\section{Introduction \label{sec:intro}}
Understanding and modeling wind speed statistics is a key for a better understanding of atmospheric turbulence and diffusion and for use in such practical applications as air quality and pollution transport modeling, estimation of wind loads on buildings, prediction of atmospheric or space probe and missile trajectory and wind power analysis. In the context of wind energy production,  it is necessary to characterize the wind in a planned wind turbine farm location, in order to calculate the optimal cut-in and cut-out speed and the likely power output of a given wind turbine (Drobinski 2012). This is done using the distribution of wind speed $M$. In particular, a key parameter to describe wind power is the average wind power produced by a wind turbine,
\begin{equation}
P_w = \frac{1}{2} \rho A C_p \bar{M}^3 K_e,
\label{eq:power_Weibull}
\end{equation}
where $A$ is the area of the flow, $\rho$ is the density of the fluid, $C_p$ is the coefficient of power of the wind turbine which measures how efficiently the wind turbine converts the energy in the wind into electricity, $K_e$ is the energy pattern, and $\bar{M}^3$ is the third moment of the distribution. In order to describe the wind energy potential independently of the turbine shapes, observed histograms are fitted to the Weibull distribution, defined as follows,
\begin{equation}
p\left(M \right) = \frac{k}{c} \left( \frac{M}{c} \right)^{k-1} \exp \left[ -
\left( \frac{M}{c} \right)^{k} \right],
\label{eq:Weibull}
\end{equation}
where $M$ is the wind speed, $k > 0$ is the shape parameter and $c > 0$ is the scale parameter of the distribution. In this case, $\bar{M^3} = c^3 \Gamma \left(1+3/k \right)$ and $K_e = \Gamma \left(1+3/k \right)/\Gamma^3 \left(1+1/k \right)$, where $\Gamma$ is the Gamma function. The Weibull distribution is thus an extremely used paradigm to model wind statistics (e.g. Justus et al. 1976, 1978, Seguro and Lambert 2000, Cook 2001, Weisser 2003, Ram\'irez and Carta 2005) because
it can be easily fit after estimating the parameters $k$ and $c$ and it produces analytical expressions for the different moments and the wind power. However, its use is purely empirical and there is a lack of physical background justifying the use of the Weibull distribution to model wind statistics. Besides, the validity of the description of wind statistics by a Weibull distribution is not clearly established. Therefore, the estimated wind potential might be systematically mis-estimated.

In this paper, we use meteorological data to show that, in some locations with particularly high wind potential but wind anisotropy, wind statistics \emph{is not well described} by the Weibull distribution, thus leading to a systematic under-estimation of wind power, up to $12\%$. We then provide physical insights for the origin of such a discrepancy. To this end, we consider wind velocity as a bivariate distribution of the two orthogonal ({\it e.g.} north-south and east-west) wind components and we compute it analytically in the simple case of normal distributions. This allows us to derive a Rayleigh-Rice super-statistical distribution, which is better suited than the Weibull distribution to describe real data in locations where the wind is not isotropic. Finally, we generalize this approach to the case of isotropic wind with non-Gaussian wind components, where the Weibull distribution better describes the meteorological data.

\section{Observations \label{sec:obs}}
\begin{figure}[t!]
\begin{center}
\includegraphics[angle=0,width=0.9 \columnwidth]{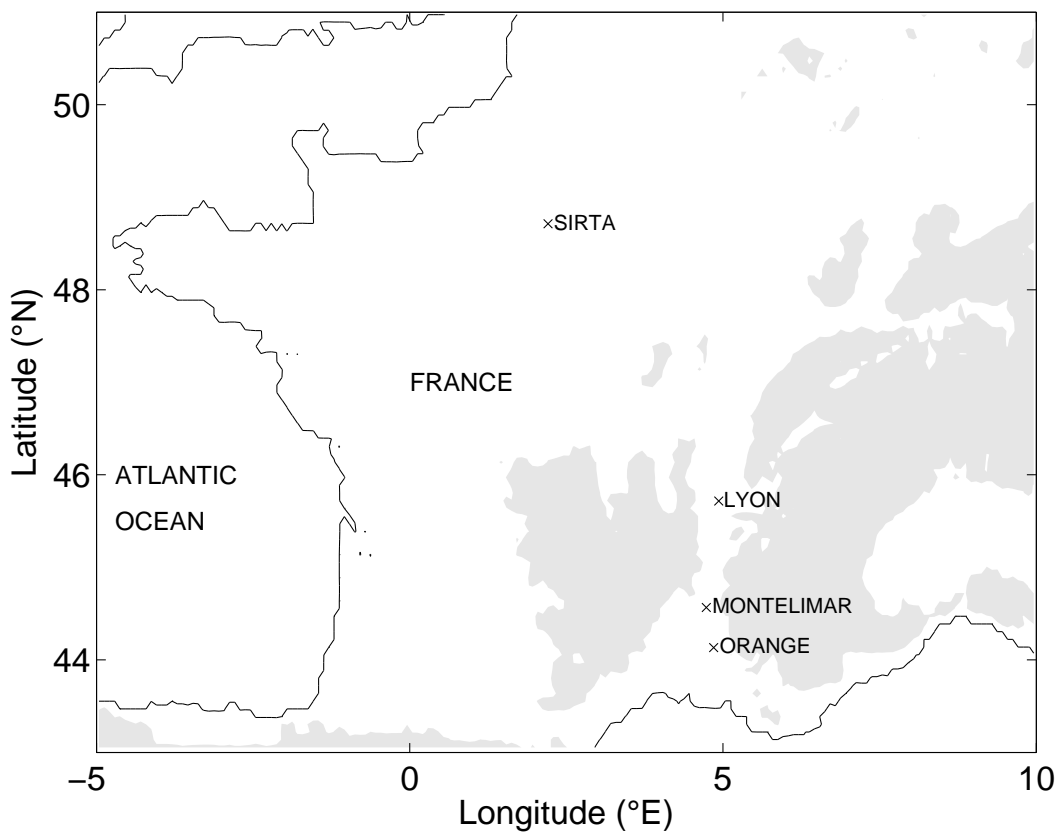}
\caption{Region of interest with the shaded area indicating a topography higher than 500~m. Locations of the surface weather stations of the SIRTA research observatory, 25~km South of Paris, and from the M\'et\'eo-France network at Lyon, Mont\'elimar and Orange are indicated with a "$\times$".}
\label{fig:topo}
\end{center}
\end{figure}

In this study, we use wind measurements collected between December, 28$^{st}$ 2006 and March 20$^{th}$ 2008 at 10~m height from 4 surface weather stations at different locations in France (Fig.~\ref{fig:topo}). Among the 4 stations, 3 are operational weather stations from the French national weather service M\'et\'eo-France and one is the research observatory (Haeffelin et al. 2005). While the wind components from the 3 operational weather stations are recorded every three hours, the SIRTA observatory data is recorded every minute. The SIRTA observatory  is located at Palaiseau, 20~km south of Paris, France. The site is in a suburban area. We use the wind measurements on a tower surrounded by different types of surface, i.e. close and distant forest, buildings and an open field sector (Drobinski et al. 2006, Barthlott et al. 2007, Fesquet et al. 2009).

The M\'et\'eo-France stations are located in a much more complex environment with elevated mountain ridges. In this region, frequent channelled wind can persist for several days. The strongest and most frequent channeled valley wind is the mistral, which blows from the north/north-west in the Rh\^one valley, separating the French Alps (highest elevation 4807 m) to the east from the Massif Central (highest elevation 1885 m) to the west by a gap 200 km long and 60 km wide (Fig.~\ref{fig:topo}) (Drobinski et al. 2005). It occurs when a synoptic northerly flow impinges on the Alpine range. As the flow experiences channeling, it is substantially accelerated and can extend offshore over horizontal ranges exceeding few hundreds of kilometers (Salameh et al. 2007, Lebeaupin-Brossier and Drobinski 2009). The mistral occurs all year long but exhibits a seasonal variability either in speed and direction, or in its spatial distribution (Mayencon 1982, Orieux and Pouget 1984, Gu\'enard et al. 2005, Gu\'enard et al. 2006). During summer, the mistral shares its occurrence with a northerly land breeze and southerly sea-breeze (e.g. Bastin and Drobinski, 2005, 2006; Drobinski et al., 2006, 2007), which can also be channelled in the nearby valleys (Bastin et al., 2005) or interact with the mistral (Bastin et al., 2006). In such region, accounting for such persistent wind systems for modeling the wind speed statistics is thus mandatory.

\begin{figure}[t!]
\begin{center}
\includegraphics[width=0.48 \columnwidth]{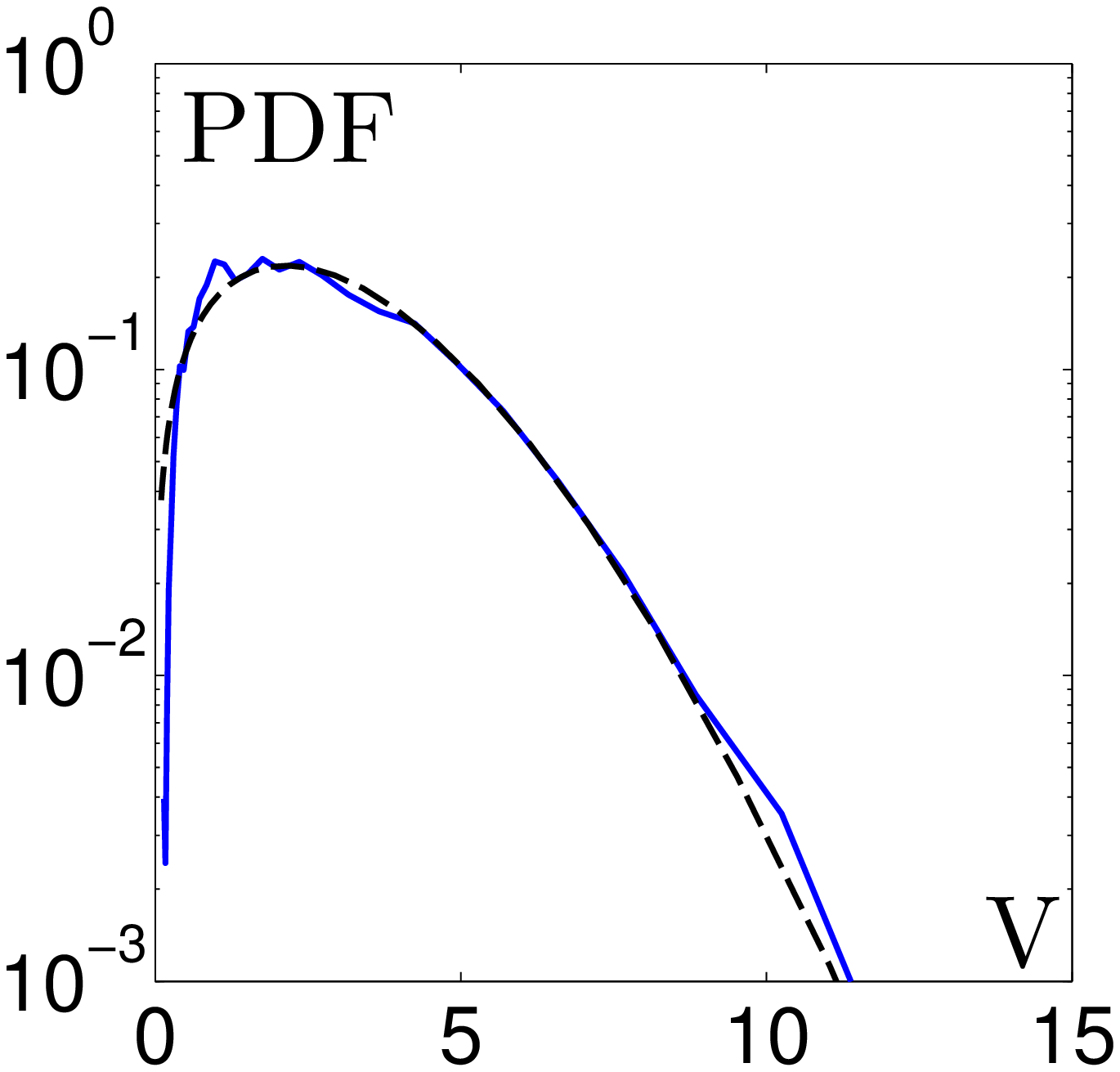}
\includegraphics[width=0.48 \columnwidth]{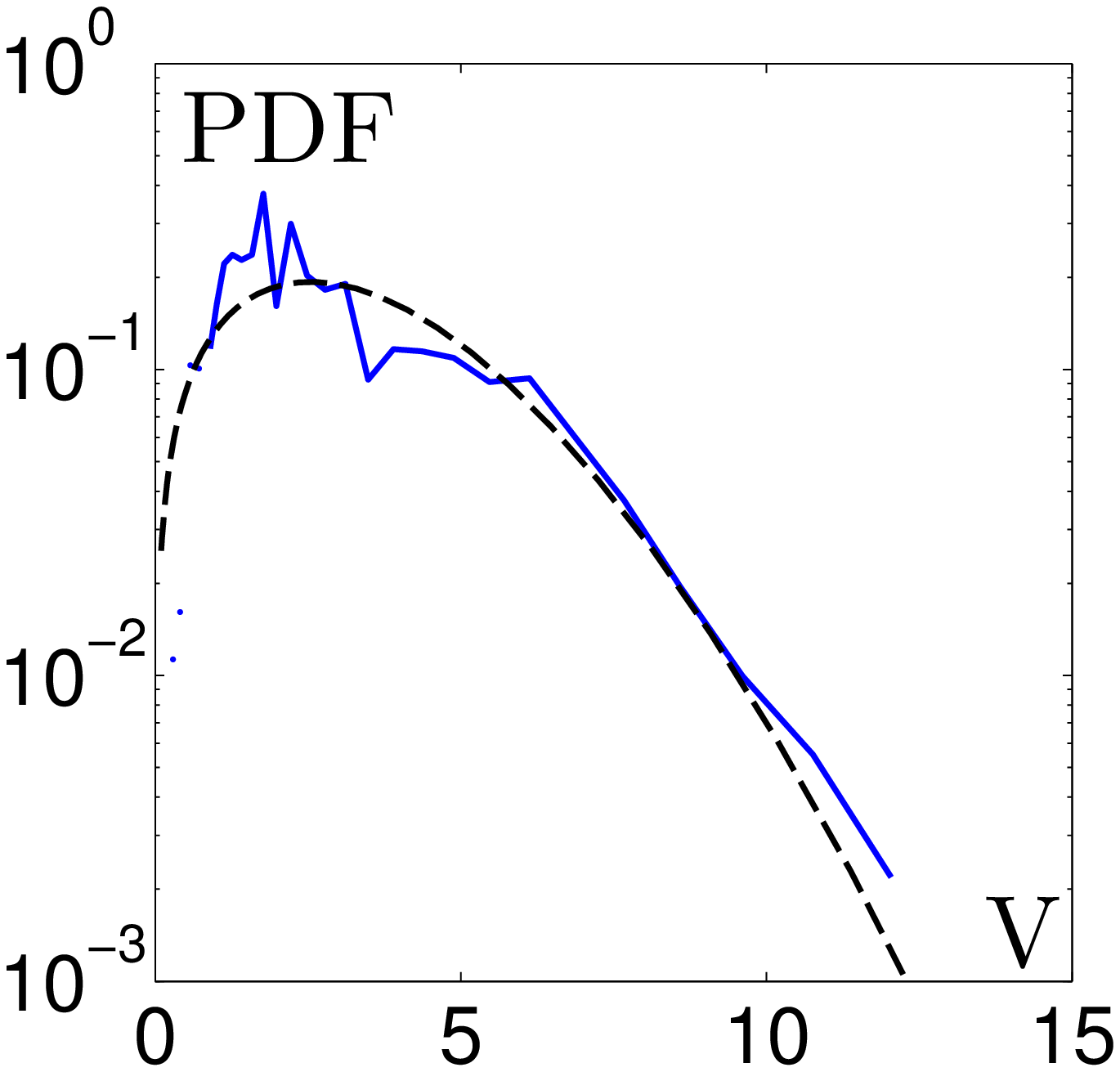}
\vspace{-2eM}\begin{flushleft}(a)\hspace{0.48\columnwidth}(b)\end{flushleft}
\includegraphics[width=0.48 \columnwidth]{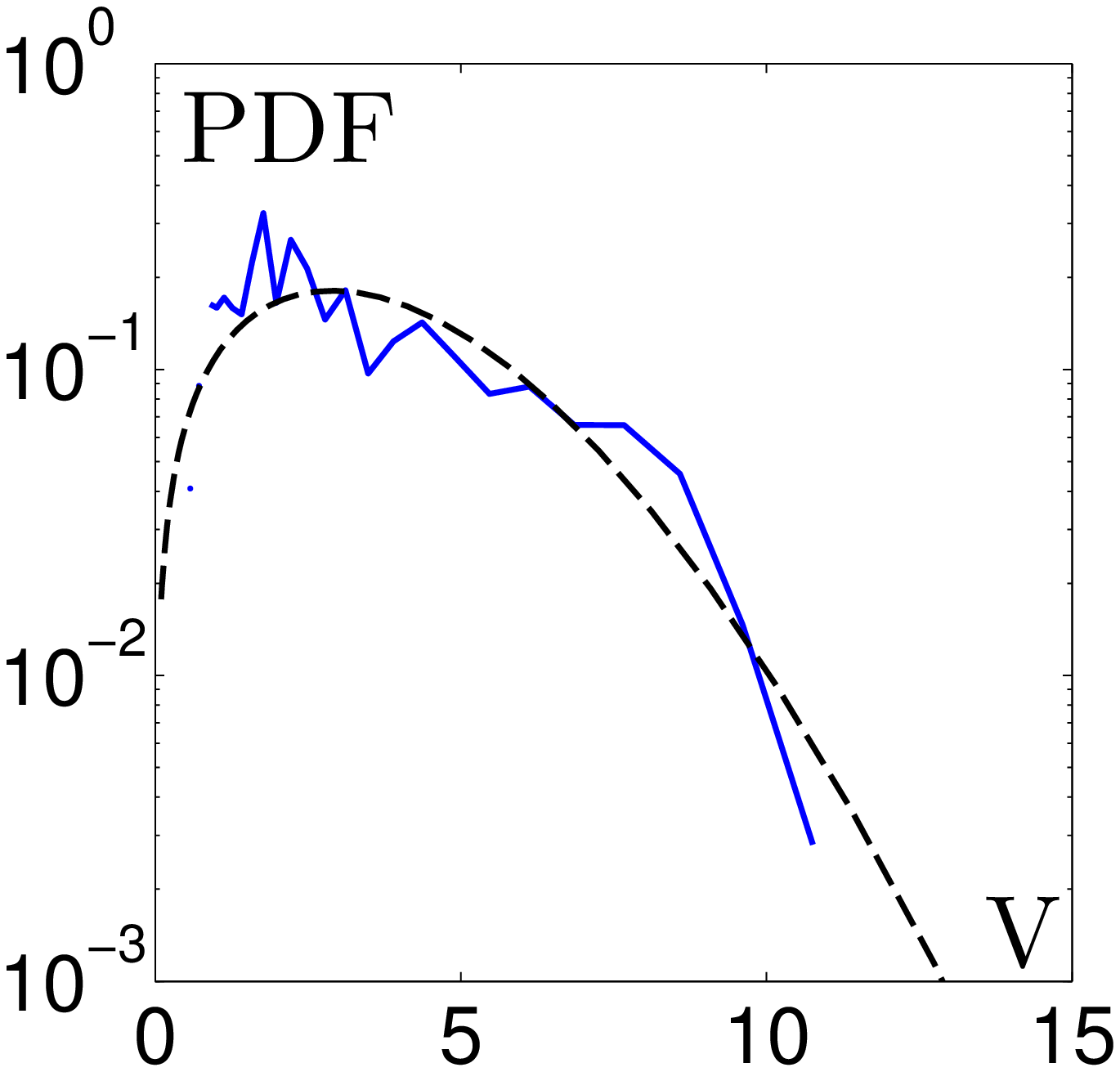}
\includegraphics[width=0.48 \columnwidth]{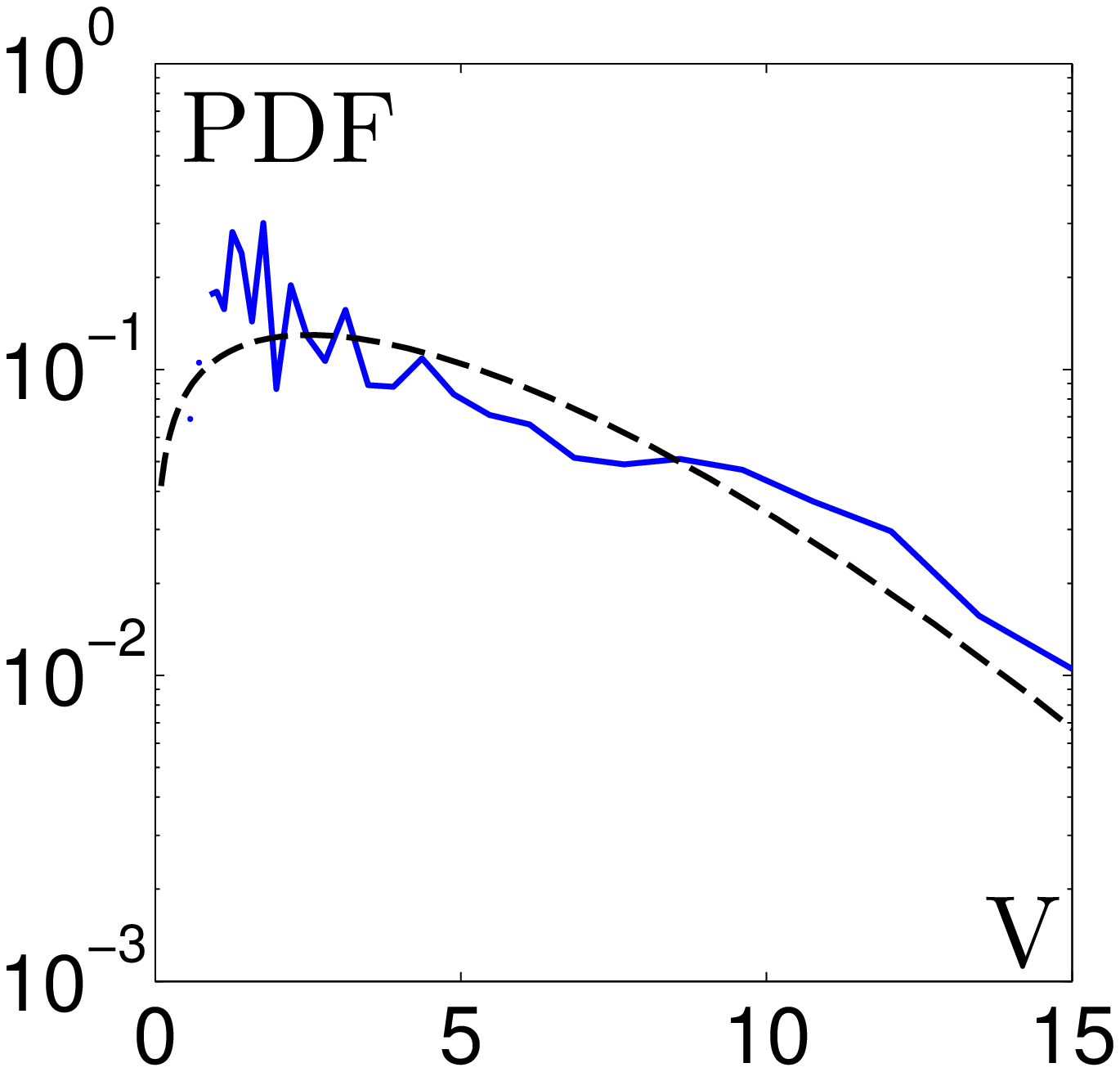}
\vspace{-2eM}\begin{flushleft}(c)\hspace{0.48\columnwidth}(d)\end{flushleft}
\caption{Probability density functions (PDF) of wind speed (blue) in  m~s$^{-1}$ and a fit by a Weibull function (dashed black) from observations made at SIRTA \leg{(a)}, Lyon \leg{(a)}, Montelimar \leg{(c)} and  Orange \leg{(d)}.}
\label{fig:wind_pdf_modulus}
\end{center}
\end{figure}

Figure~\ref{fig:wind_pdf_modulus} shows the probability density functions (PDF) of the wind speed at the SIRTA observatory, Lyon, Mont\'elimar and Orange and their fits by a Weibull distribution.
First, while the Weibull distribution is an accurate statistical model for the wind speed distribution at SIRTA (Fig.~\ref{fig:wind_pdf_modulus}a), it poorly describes wind distributions measured at the 3 weather stations located in complex terrain (Fig.~\ref{fig:wind_pdf_modulus}b-d). Table~\ref{tab:comp_obs_pdf_vs_model_pdf} compares the wind power computed from the wind speed observations and that is estimated from the Weibull distribution best fit using Eq.~(\ref{eq:power_Weibull}). The parameters $k$ and $c$ are also indicated. At the SIRTA observatory, the wind power estimated from Eq.~(\ref{eq:power_Weibull}) is accurate up to $2.5$~\%. It is obtained for a value $k=1.71$. The Weibull distribution describes fairly well the wind distribution at SIRTA, which we stress, is located in a fairly flat terrain. Yet, the difference between the data and the fit is not random fluctuations:
there are small systematic differences occurring at large ($M> 8$~m.s$^{-1}$) and at low ($M< 2$~m.s$^{-1}$) wind speed values.
\begin{table}
\begin{indented}
\item[]\caption{\label{tab:comp_obs_pdf_vs_model_pdf} Comparison of the power computed from the measured data and out of the different fits (Weibull, Elliptical and Rayleigh-Rice distributions; see text for definitions) using typical values $Cp=0.45$, $\rho=1.2$~kg$/$m$^3$ and $A=7.9\times10^3$~m$^2$. The quantities $k$ and $c$ are the estimated Weilbull distribution parameters, $\sigma_u$ and $\sigma_v$ the estimated standard deviations of the 2 wind components of the Elliptical distribution and $\alpha$ is the estimated regime parameter (see text for definition).}
 \begin{tabular}{ccccc}
\br
Stations                     & \multicolumn{4}{c}{Power $\times(10^2 kW)$} \\
\cline{2-5}
                             &  Data  &   Weibull   &         Elliptic&    Rayleigh-Rice    \\
                             &        &  [$k$;$c$]  &  [$\sigma_u$;$\sigma_v$] & [$\alpha$] \\
\mr
SIRTA                        &  1.63  &     1.59    &            1.48     &   1.13          \\
                             &        & [1.71;3.58] &         [2.64;2.64]      &[0.05]            \\
Lyon                         &  2.26  &     2.22    &            2.17
&    2.26    \\
                             &        & [1.76;4.11] &         [2.33;3.54]
&   [0.297]  \\
Montelimar                   &  2.75  &     2.55    &            2.53
&    2.74    \\
                             &        & [1.82;4.48] &         [2.65;3.73]
&   [0.230]  \\
Orange                       &  7.71  &     6.78    &            6.75
&    7.59    \\
                             &        & [1.45;5.69] &         [2.93;5.53]
&   [0.510]  \\
\br
 \end{tabular}
\end{indented}
\end{table}

\begin{figure}[ht!]
\begin{center}
\includegraphics[width=0.48 \columnwidth]{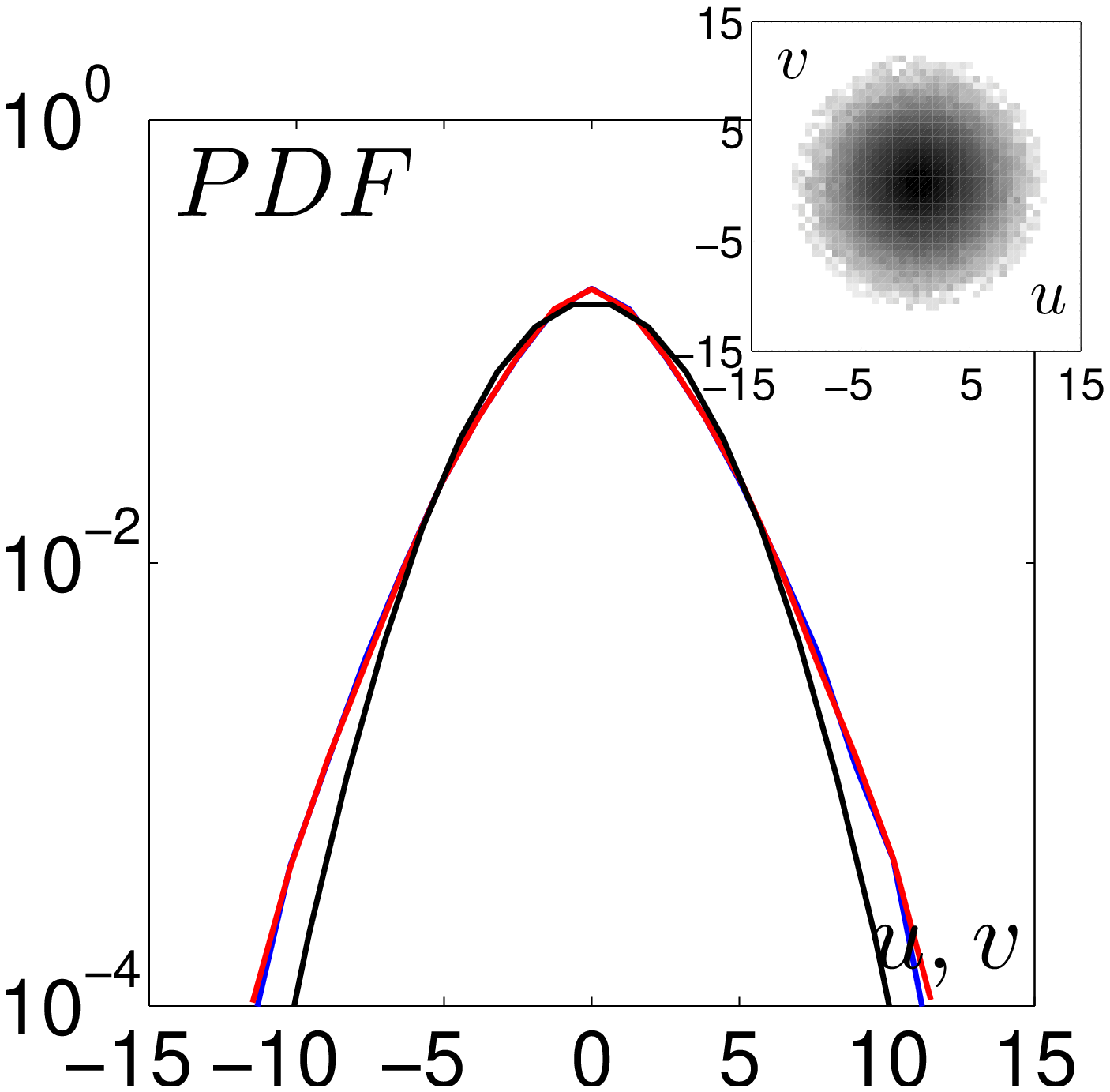}
\includegraphics[width=0.48 \columnwidth]{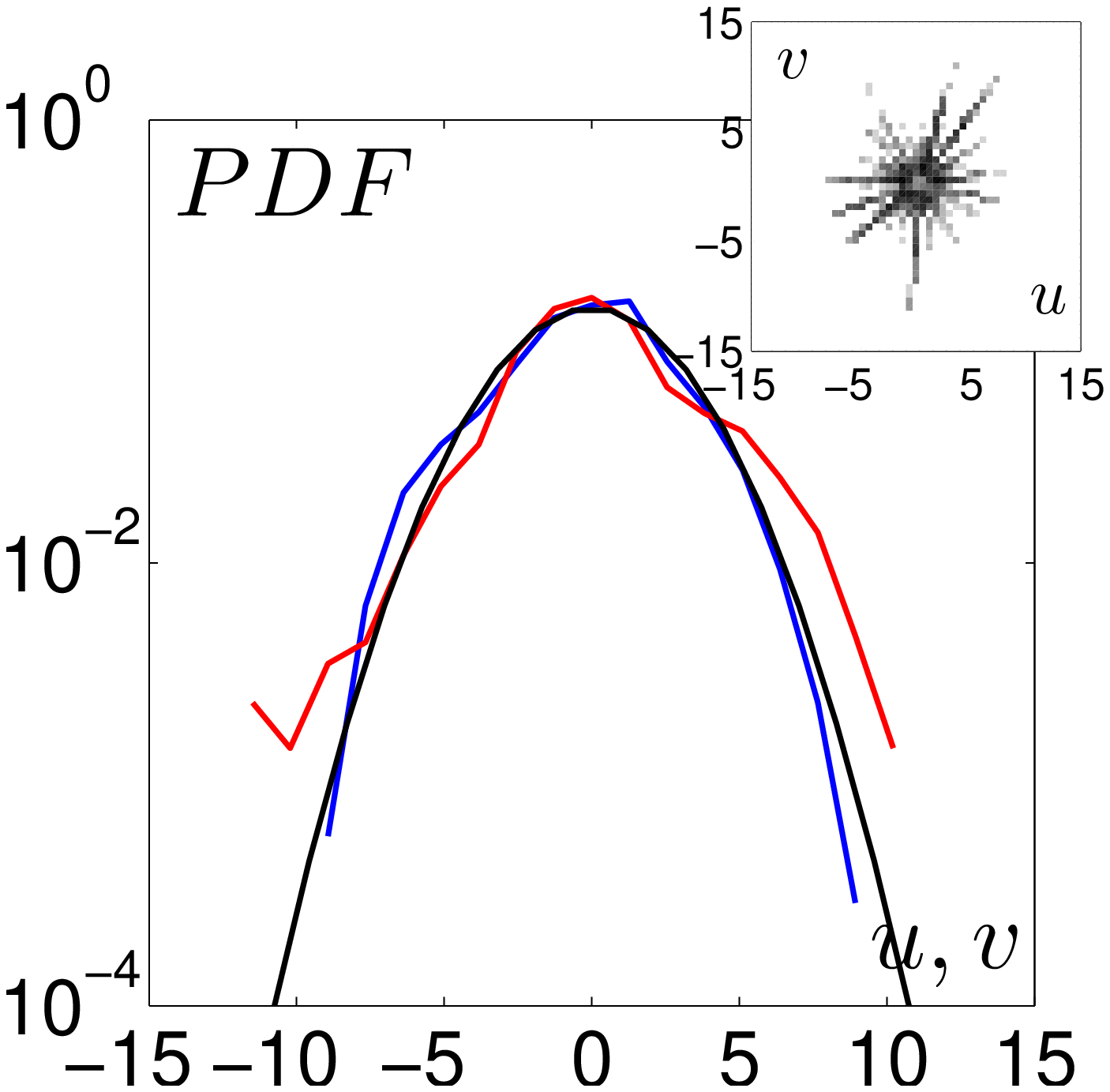}
\vspace{-2eM}\begin{flushleft}(a)\hspace{0.48\columnwidth}(b)\end{flushleft}
\includegraphics[width=0.48 \columnwidth]{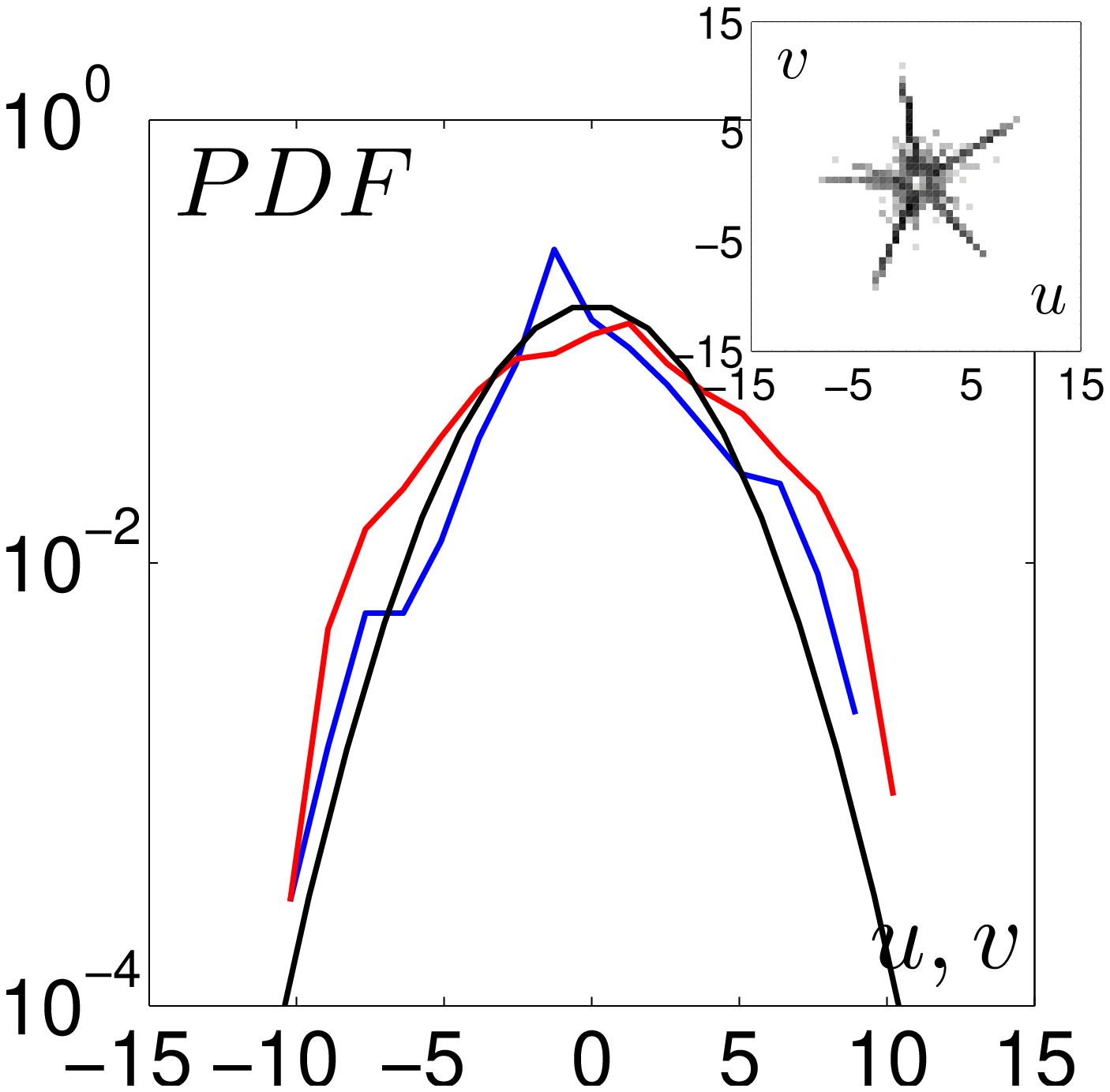}
\includegraphics[width=0.48 \columnwidth]{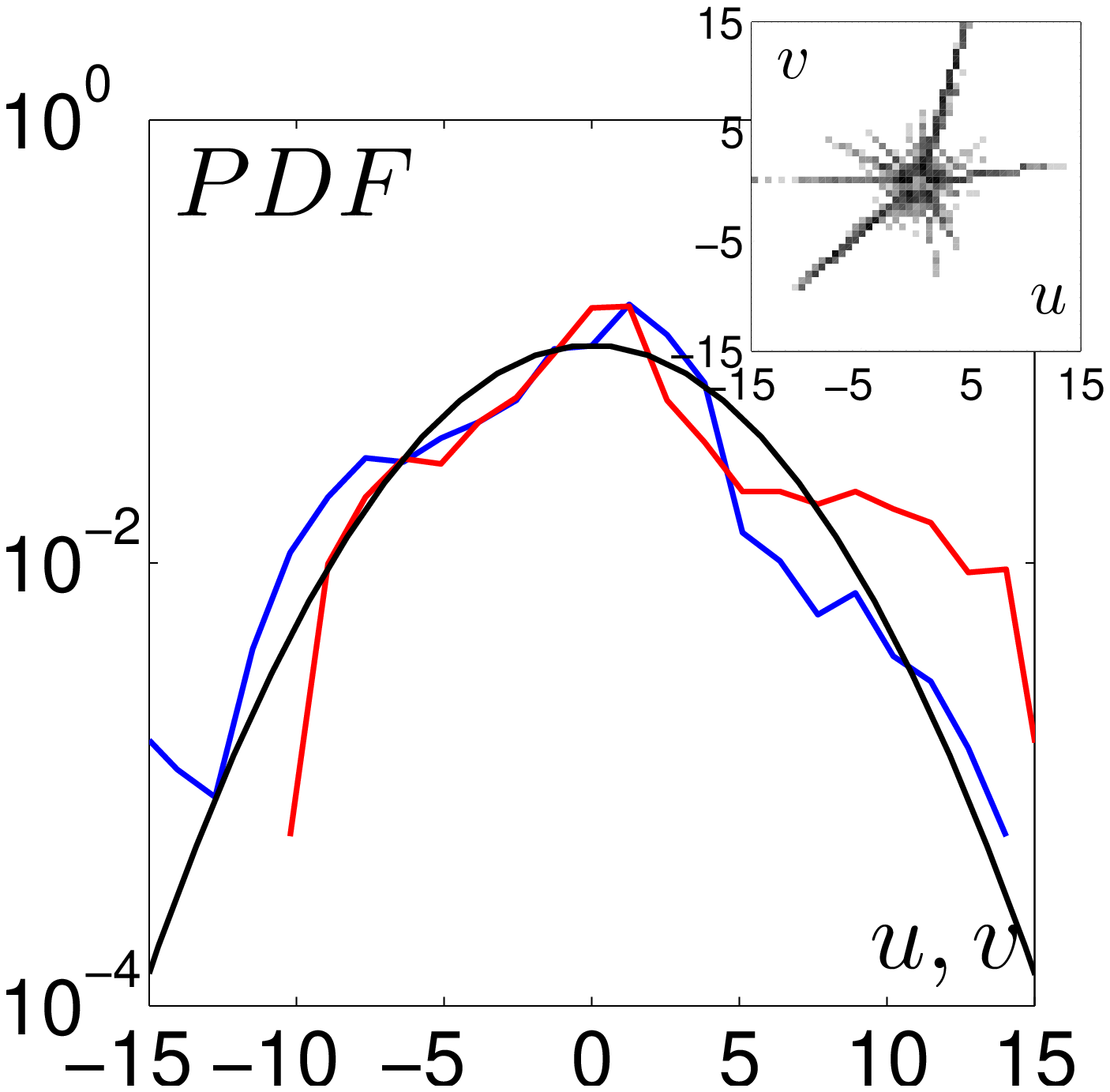}
\vspace{-2eM}\begin{flushleft}(c)\hspace{0.48\columnwidth}(d)\end{flushleft}
\caption{Distributions of north-south (red) and East-West (blue) wind components in m~s$^{-1}$ and a Gaussian function (dashed black) from observations made at (\leg{a}) SIRTA, (\leg{b}) Lyon, (\leg{c}) Montelimar and (\leg{d}) Orange. \leg{Insets :} Joint distributions of wind components.}
\label{fig:wind_pdf_components}
\end{center}
\end{figure}

In contrast, at Mont\'elimar and Orange, Weibull fit quality drops (Fig.~\ref{fig:wind_pdf_modulus}c-d), and systematically underestimates the right tail of the distributions. As a result, the power is systematically underestimated by $7.5$~\% and $12.8$~\% (Table~\ref{tab:comp_obs_pdf_vs_model_pdf}). To the best of our knowledge, this systematic discrepancy has not been reported in the existing literature, although it is a major issue for wind power resource evaluation. At Lyon, Mont\'elimar and Orange, the observed distributions display complex shapes instead of a smooth and regular shape as would be expected when fitting with a Weibull distribution. As opposed to the regular shape of wind speed and the surrounding topography at SIRTA, the complexity of the distributions in the Rh\^one valley might originate in the complexity of the surrounding topography.
For a deeper insight of the differences between observed wind speed distributions and the commonly used Weibull distribution, the distribution of the wind components are now investigated (Fig.~\ref{fig:wind_pdf_components}). At the SIRTA observatory (Fig.~\ref{fig:wind_pdf_components}a), the zonal and meridional wind components have the same symmetrical distributions with zero mean and equal width: the wind is isotropic. In addition, the distributions of wind components have larger tails than Gaussian.

At Lyon (Fig.~\ref{fig:wind_pdf_components}b), at the entrance of the Rh\^one valley, the PDF of the zonal wind component is still symmetrical with almost zero mean and has a Gaussian shape. The PDF of the meridional wind component is fairly symmetrical for values below 5~m~s$^{-1}$ but is skewed for the strongest values. Also the width of the meridional wind PDF is larger than that of the zonal wind. At Mont\'elimar and Orange, none of the wind component PDFs look like Gaussian distribution and the meridional wind PDF become significantly larger than the PDF of the zonal wind. This shows evidence of the channelled nature of the wind by the north-south oriented Rh\^one valley (Salameh et al. 2009). For low winds, the PDFs of the zonal and meridional components are more similar, evidencing a less efficient impact of the valley flanks to orientate the wind.

The two wind components may however be correlated. The insets of Fig.~\ref{fig:wind_pdf_components} show the joint distribution of the two wind components at SIRTA observatory, Lyon, Mont\'elimar and Orange. At the SIRTA observatory, the rather flat terrain produces an isotropic wind statistics ({\it i.e.} the wind blows in all directions) with no correlation between the two components ({\it i.e.} the joint distribution displays a circular shape). Conversely, the joint distributions at Lyon, Mont\'elimar and Orange display strongly correlated wind components in specific directions. The main directions correspond to the north-south orientation of the Rh\^one valley.

To conclude on the observations, it appears that at SIRTA and Lyon, while the Weibull function fits better wind speed distributions, the wind components are less anisotropic. Therefore, it seems that isotropy is crucial for the Weibull description to be valid. In the following, despite the fact that none of the observed distributions is Gaussian, we shall focus on anisotropic bivariate distribution to understand the effect of anisotropy on the resulting wind speed distribution.

\section{From Gaussian to super-statistical Gaussian wind components}

\subsection{The Rayleigh distribution}
\begin{figure}[b!]
\begin{center}
\includegraphics[width=0.48 \textwidth]{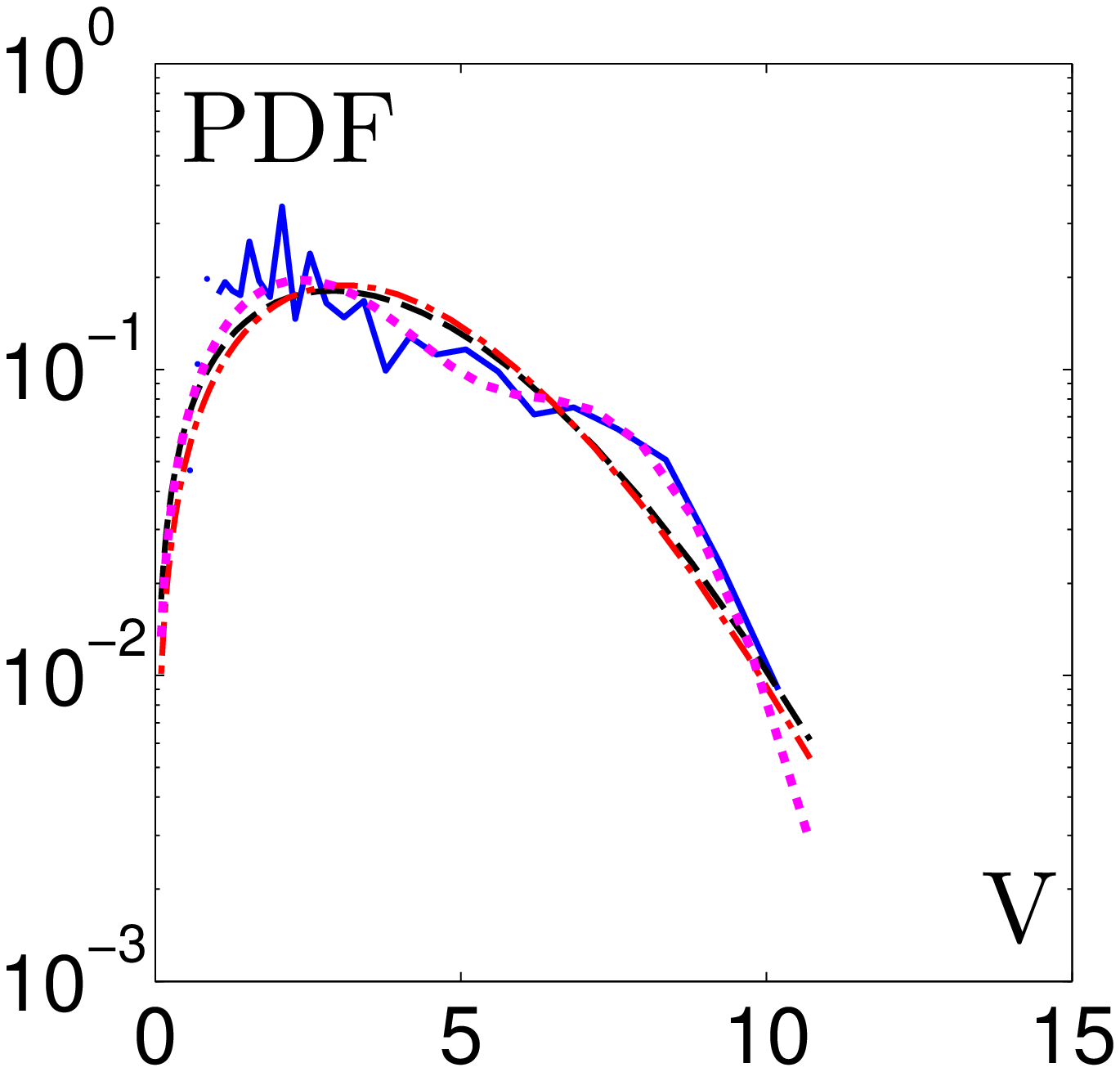}
\includegraphics[width=0.48 \textwidth]{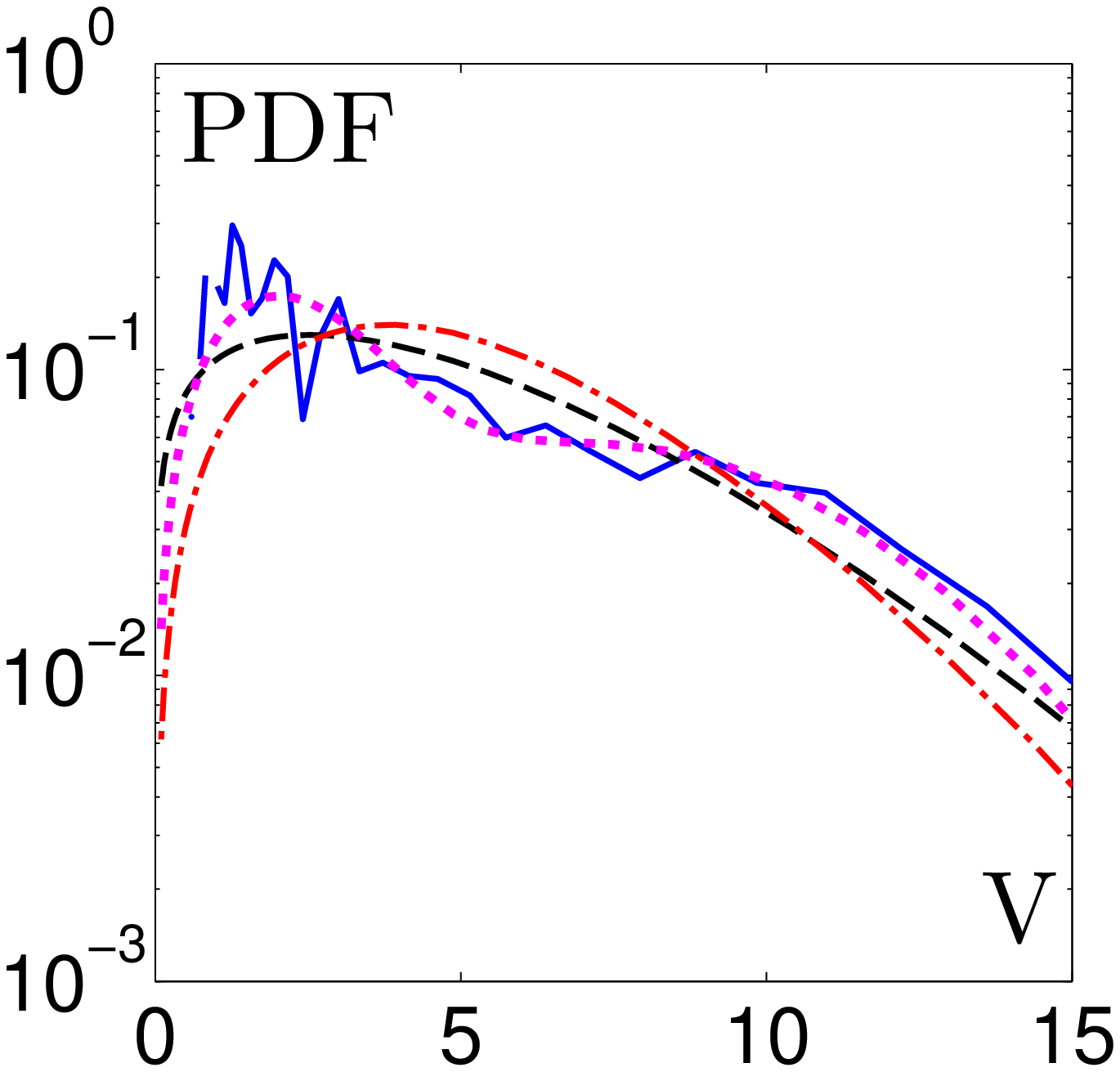}
\vspace{-2eM}\begin{flushleft}(a)\hspace{0.48\columnwidth}(b)\end{flushleft}
\caption{Distributions of wind speed (blue) in  m~s$^{-1}$ and a fit by a Weibull function (dashed black), Elliptic function (dash-dot red), and weighted Rayleigh-Rice function (dotted magenta) from observations made at (\leg{a}) Montelimar and (\leg{b}) Orange.}
\label{fig:wind_pdf_modulus_3}
\end{center}
\end{figure}

The very first approach to model anisotropy is to consider a bivariate distribution of Gaussian wind components with means $\mu_u$ and $\mu_v $, and variances $\sigma_u$ and $\sigma_v$ and finite cross-correlation $\rho$. After a change of variable, $u\rightarrow u'$, $v\rightarrow v'$, the resulting joint distribution is a bivariate elliptical distribution with variances $\sigma_u'$ and $\sigma_v'$ without correlation. In the case $\mu_u' = \mu_v' = 0$ (Crutcher and Baer 1962),
\begin{eqnarray}
\begin{array}{l}
\displaystyle p(u',v';\sigma_u'^2,\sigma_v'^2) = \frac{1}{2 \pi \sigma_u' \sigma_v'}
\exp\left( -\frac{u'^2}{2\sigma_u'^2}+\frac{v'^2}{2\sigma_v'^2} \right).
\end{array}
\label{eq:bivariate_gauss_pdf_indep_anisotrop}
\end{eqnarray}
Applying the usual transformation from Cartesian to polar coordinates ($M$,$\phi$), and integrating over the angle $\phi$, the joint PDF for $M$ and $\phi$ is (Chew and Boyce 1962):
\begin{equation}
p_M(M;\sigma_u'^2,\sigma_v'^2) = \frac{M}{\sigma_u' \sigma_v'}  \exp\left( -aM^2
\right)I_0\left( bM^2 \right),
\label{eq:wind_speed_pdf_anisotrop}
\end{equation}
with $a = (\sigma_u'^2+\sigma_v'^2)/(2\sigma_u' \sigma_v')^2$ and $b = (\sigma_u^2-\sigma_v^2)/(2\sigma_u \sigma_v)^2$, and  where $I_0(x)$ is the modified Bessel function of the first kind and zero order. When the variance are equal $b=0$, so $I_0(bM^2)=1$ and Eq.~(\ref{eq:wind_speed_pdf_anisotrop}) is the Weibull distribution with parameter $k=2$, which is the Rayleigh distribution. When comparing the elliptical normal joint distribution with data, (Table~\ref{tab:comp_obs_pdf_vs_model_pdf} and Fig.~\ref{fig:wind_pdf_modulus_3}), it is worse than a fit by the Weibull distribution.
Despite the fact taking account of anisotropy of wind components modifies the tail of the distributions,
the Gaussian approximation is too crude to model wind speed components. However, using Gaussian distributions is
a powerful approach because analytic calculation can be performed.

\subsection{The Rice distribution}
As a second order approximation and for the sake of simplicity, we now consider the case of elliptic distributions with non zero mean $\mu_u$ and $\mu_v$ and equal variances $\sigma_u'=\sigma_v'=\sigma$. equal variances for the two wind components. Thus, we obtain
\begin{equation}
p_M(M;\mu,\sigma^2) =  \frac{M}{\sigma^2}
 \exp\left( -\frac{M^2+\mu^2}{2\sigma^2} \right) I_0 \left(
\frac{M\mu}{\sigma^2} \right),
\label{eq:wind_speed_pdf_mean_wind}
\end{equation}
with $\mu = \sqrt{\mu_u^2+\mu_v^2}$. The distribution in Eq.~(\ref{eq:wind_speed_pdf_mean_wind}) is called the Rice distribution and $I_0$ is the $0^{\rm th}$ order modified Bessel function of the first kind.

\subsection{The Rayleigh-Rice distribution}
The key assumption we do in the following is that, in the Rh\^one valley area, there are two regimes: (i) random flow: the wind components have zero means and same variances, and the Rayleigh distribution describes well the wind speed statistics; (ii) channeled flow: the wind components have different means, and the Rice distribution describes well the wind speed statistics. The resulting distribution is a sum of the distribution (i) and (ii) conditioned to the absence and presence of the channeled flow occurrence. In other words, in the long-term, the wind system is described as the superposition of different local dynamics at different intervals that has been coined by Beck and Cohen (2003) as superstatistics or statistics of statistics. This yields, for one wind component, for instance the zonal wind, $u$ (similar expression would arise for the meridional component, $v$):
\begin{equation}
\displaystyle p(u) = \displaystyle \int \alpha_u(\mu)p(u)d\mu.
\label{eq:superstatistics}
 \end{equation}
In the case of the two regimes scenario, $\alpha_u(\mu)$ is bimodal,
\begin{equation}
\displaystyle \alpha_u(\mu)  = \displaystyle \left( 1 - \alpha \right)
\delta(\mu) + \alpha \delta(\mu - \mu_u),
\label{eq:alpha}
\end{equation}
where $\delta$ is the Dirac function and $\alpha$ is here the weight corresponding to channeled flow events occurrence. The PDF of $M=\sqrt{u^2+v^2}$, is thus a combination of a Rayleigh distribution (corresponding to $\mu_u = \mu_v = 0$) and a Rice distribution. Such a method has been used to include in wind statistics modeling the contribution of zero wind speed (Tackle and Brown 1978, Tuller and Brett 1983). As a result we obtain the Rayleigh-Rice distribution:
\begin{equation}
\fl
p_M(M;\mu,\sigma^2,\alpha) = \frac{M}{Z}  \exp\left( -\frac{M^2}{2\sigma^2}
\right) \left(\alpha + (1-\alpha) \exp\left( -\frac{\mu^2}{2\sigma^2} \right) I_0 \left(
\frac{M\mu}{\sigma^2} \right)\right),
\label{eq:rayleigh_rice}
\end{equation}
where $Z = \int p_M(M;\mu,\sigma^2,\alpha)dM$ is the normalization factor.

When comparing to data, the weighted Rayleigh-Rice function provides a much better fit (Fig.~\ref{fig:wind_pdf_modulus_3}) and estimates much better the wind power (Table~\ref{tab:comp_obs_pdf_vs_model_pdf}) than
the Weibull distribution. We indeed obtain less than $1.5\%$ relative error for all locations and even less than $0.3\%$ in Mont\'elimar. Table~\ref{tab:comp_obs_pdf_vs_model_pdf} shows that the weighted Rayleigh-Rice function gives by far the best estimate of the wind power at Lyon, Mont\'elimar and Orange. The values of $\alpha$ correspond very accurately to the probability of occurrence of the dominating  mistral situations (Orieux and Pouget 1984). Thus, taking account of the existence of regimes in regions where mountain environments is a fruitful approach to quantify wind speed statistics and to estimate wind power potential.

\section{Non-Gaussian distributions}
\subsection{Super-statistics}
Figure~\ref{fig:wind_pdf_components} shows that at SIRTA, where the environment is much less complex than at the 3 other weather stations, the PDFs of the two wind components are not accurately predicted by a normal distribution and displays heavy tails. Such heavy tails illustrate the departure of the wind component PDFs from the normal distribution model. We show here that super-statistics can help for the understanding the non-Gaussian shape. Since both wind components at the SIRTA station have very close statistical properties (Fig.~\ref{fig:wind_pdf_components}a), for the sake of concision, we will focus on one component, $u$, throughout this section.

\begin{figure}[b!]
\begin{center}
\includegraphics[width=0.48 \columnwidth]{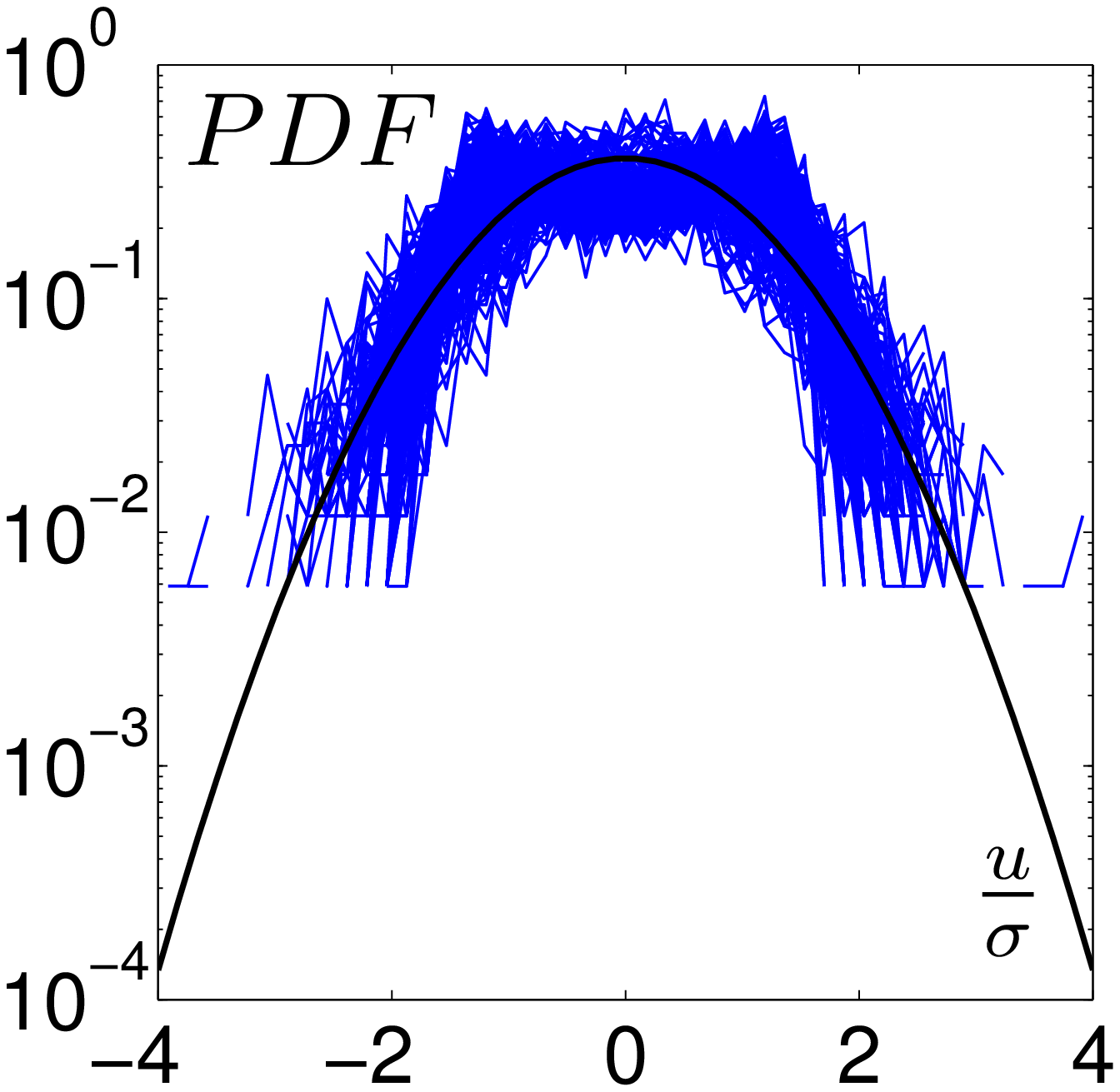}
\includegraphics[width=0.48 \columnwidth]{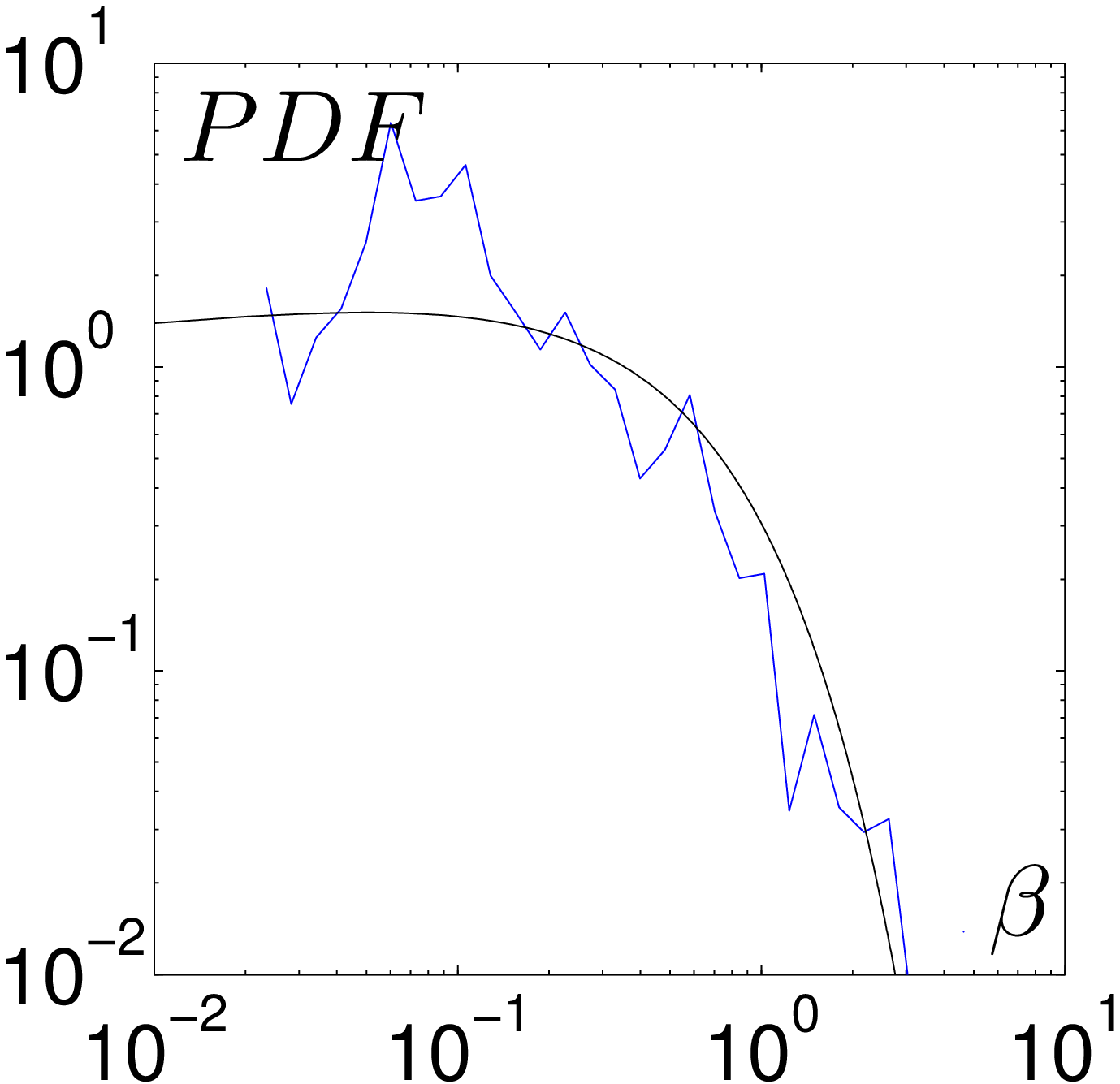}
\vspace{-2eM}\begin{flushleft}(a)\hspace{0.48\columnwidth}(b)\end{flushleft}
\includegraphics[width=0.48 \columnwidth]{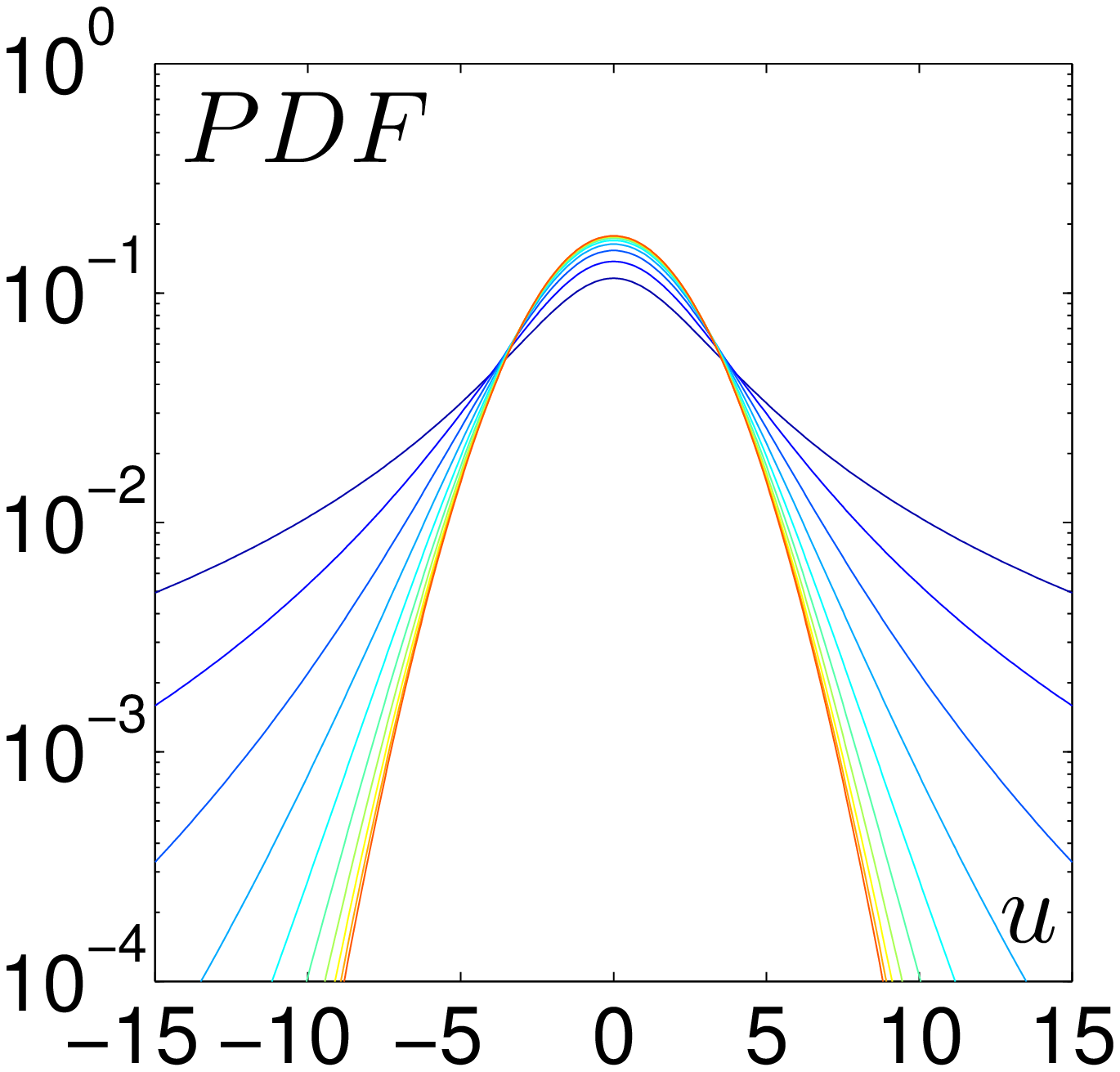}
\includegraphics[width=0.48 \columnwidth]{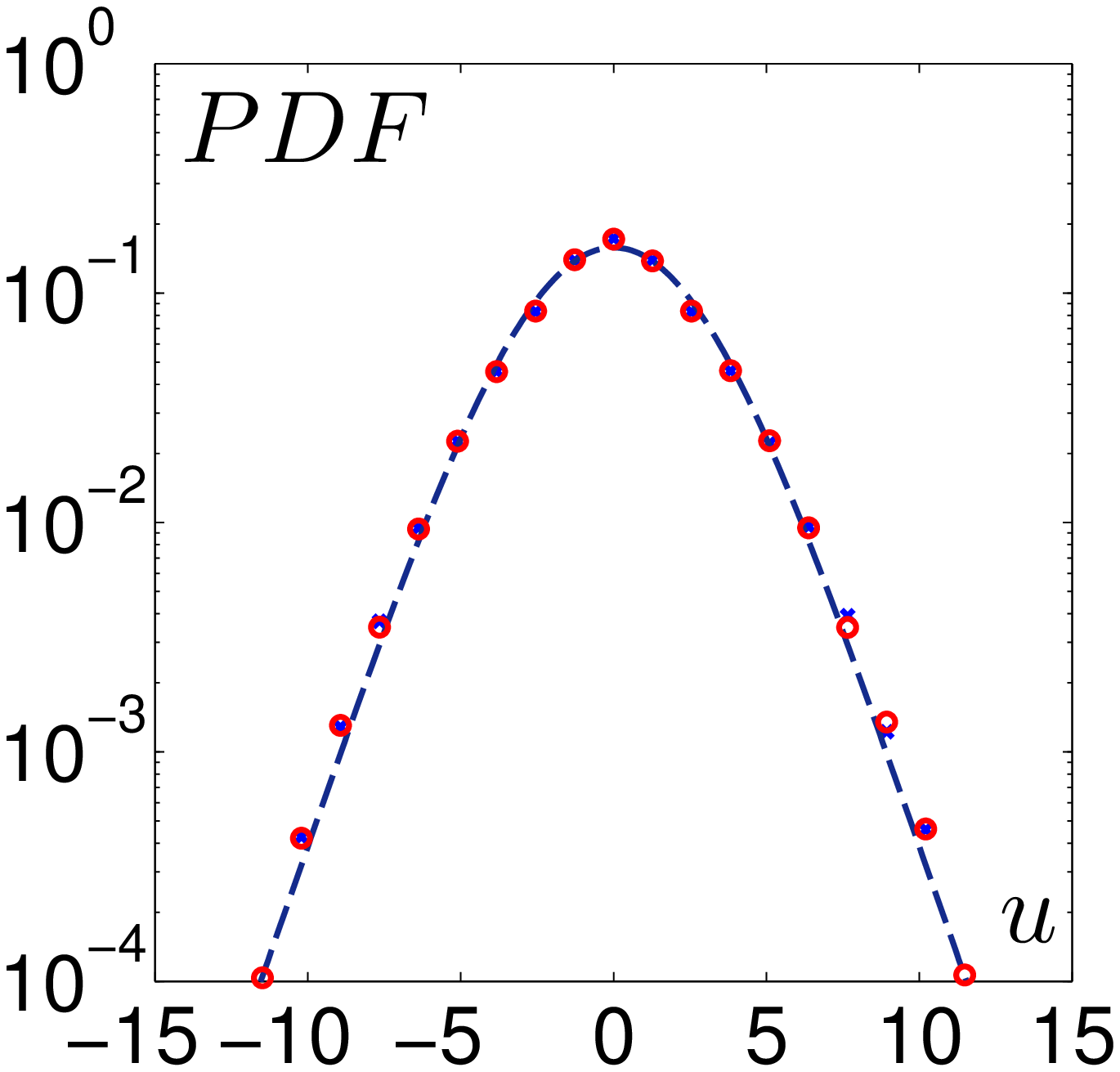}
\vspace{-2eM}\begin{flushleft}(c)\hspace{0.48\columnwidth}(d)\end{flushleft}
\caption{\leg{(a):} Distributions of short samples of wind component speed $u$, rescaled by their standard deviation $\sigma$ computed over 4~hr time periods. Black line indicates a Gaussian function of standard deviation $1$. \leg{(b):} Distribution of each sample standard deviation $\beta=\sigma^{-2}$ and its fit by a Gamma distribution. \leg{(c):} Superstatistical distribution corresponding to Eq.~(\ref{eq:superstatistics_4}). Color codes for varying parameters $m$ and $n$. \leg{(d):} Wind speed component distributions from the SIRTA observatory measurements (\textcolor{red}{$\bigcirc$}, $u$, and \textcolor{blue}{$\times$}, $v$) and their fit by the superstatistical distribution corresponding to Eq.~(\ref{eq:superstatistics_4}) with parameters, $m=9$, $n=0.019$ (black dashed line).}
\label{fig:SIRTA_superstat}
\end{center}
\end{figure}

The large tails of the wind component distributions at SIRTA (Fig.~\ref{fig:wind_pdf_components}a) originate in the transient nature of the wind. Meteorological conditions indeed change on several time-scales (day, anticyclones duration, season). For instance, the strongest wind being typically recorded in the winter, whereas in summer, long-lasting anticyclonic conditions give lower wind speeds. This induces a change of the statistical properties (e.g. wind component variance) on several entangled time scales. Here, we model this by assuming Gaussian shape of wind distribution, but with changing standard deviation, $\sigma$. The super-statistics of the wind components can be derived as follows (Rizzo and Rapisarda 2004):
\begin{equation}
\displaystyle p(u)  = \displaystyle \int f(\beta)p(u;\beta)d\beta,
\label{eq:superstatistics_2}
\end{equation}
where $f$ is the probability distribution of the fluctuating variable $\beta=1/\sigma^2$. Figure~\ref{fig:SIRTA_superstat}a shows the distributions of short samples of wind component speed $u$ measured at the SIRTA observatory, rescaled by their standard deviation, $\sigma$, computed over 4~hr time periods.
We see that the distributions of wind component velocity acquired during short time samples is well rescaled by its standard deviation. We can therefore assume the wind component distributions to be
\begin{equation}
p(u;\beta) = \sqrt{\frac{\beta}{\pi}}\exp\left(-\beta u^2\right).
\label{eq:superstatistics_3}
\end{equation}
Figure~\ref{fig:SIRTA_superstat}b shows the distribution of each sample standard deviation $\beta=\sigma^{-2}$ and its fit by a Gamma distribution. We see that the distribution of $\beta$ is consistent with a Gamma distribution, of parameters $m$ and $n$:
\begin{equation}
\displaystyle f(\beta)  = \displaystyle \frac{1}{\Gamma(m+1)n} \left(\frac{\beta}{n}\right)^m \exp \left(-\frac{\beta}{n}\right),
\label{eq:superstatistics_gamma}
\end{equation}
where $\Gamma$ is the Gamma function. Hence, combining it to Eq.~(\ref{eq:superstatistics_2}), we obtain
\begin{equation}
\displaystyle p(u)  = \displaystyle \frac{\sqrt{n/2}}{B(\frac{1}{2},m-1)} \left( 1 + n u^2 \right)^{-m},
\label{eq:superstatistics_4}
\end{equation}
where $B$ is the Euler integral of the first kind. This distribution is called the generalized Boltzmann factor. It has been obtained when attempting to generalize the entropy definition (Tsallis 1988, Beck and Cohen 2003). The distributions modeled with Eq.~(\ref{eq:superstatistics_4}) display heavy power law tails reminiscent of rare events (Fig.~\ref{fig:SIRTA_superstat}c). Unfortunately, our data has not enough statistics to observe such rare events and we can not compare the tails of the distribution (Fig.~\ref{fig:SIRTA_superstat}d). Yet, note that the central part of the observed distributions matches very well the distributions of Eq.~(\ref{eq:superstatistics_4}) (Fig.~\ref{fig:SIRTA_superstat}d).

\begin{figure}[t!]
\begin{center}
\includegraphics[width=0.48 \columnwidth]{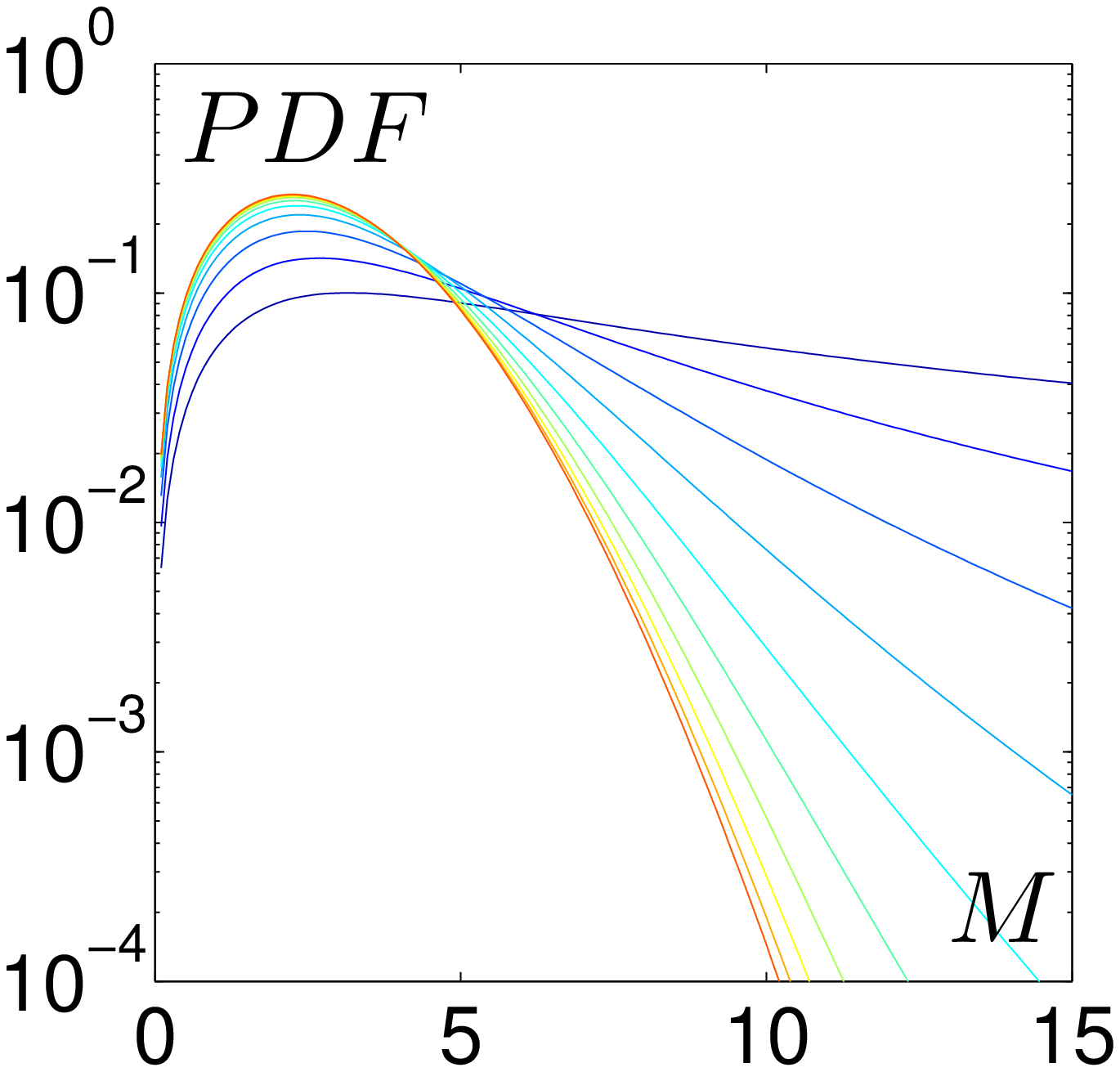}
\includegraphics[width=0.48 \columnwidth]{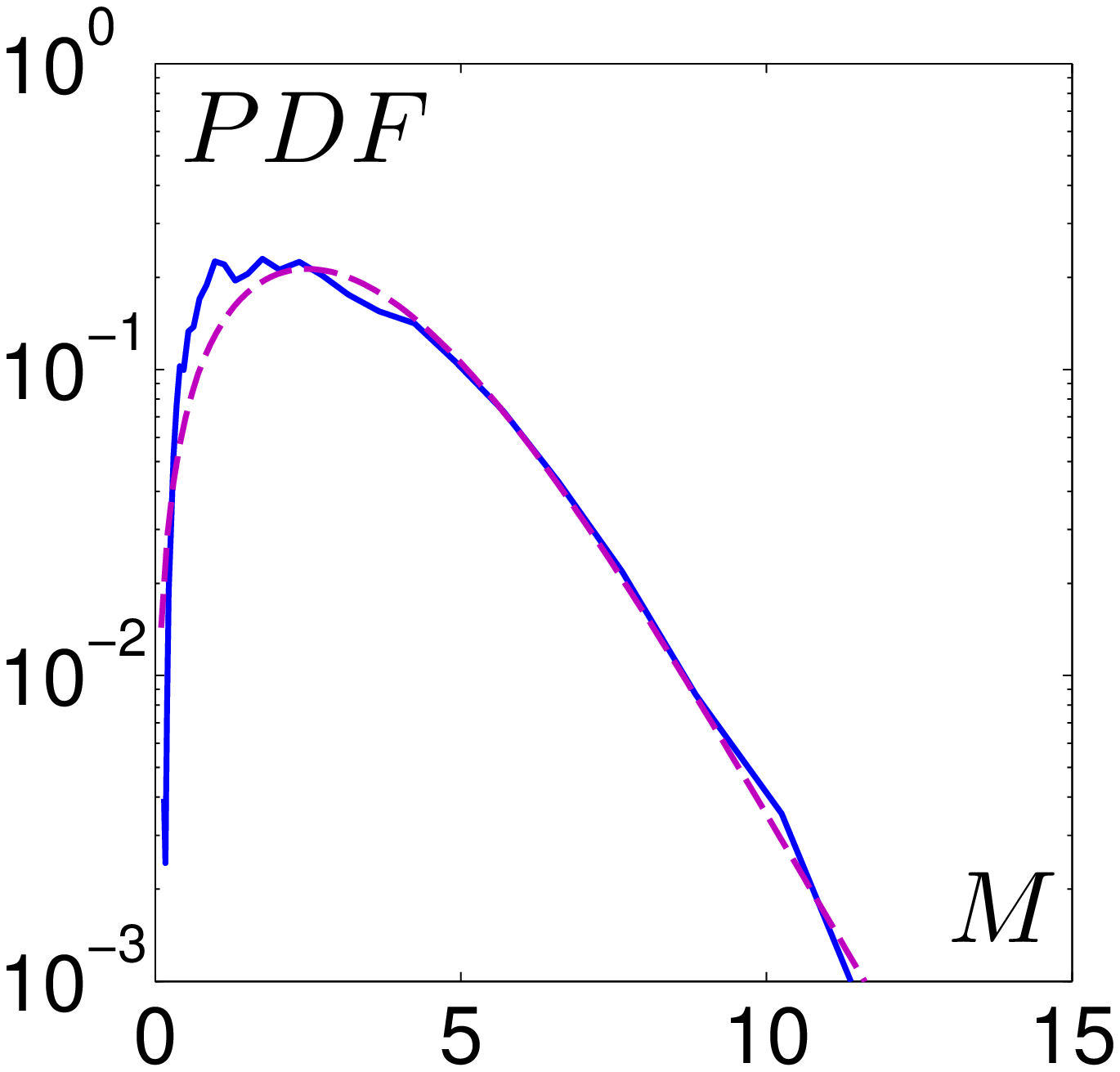}
\vspace{-2eM}\begin{flushleft}(a)\hspace{0.48\columnwidth}(b)\end{flushleft}
\caption{\leg{(a):} Superstatistical distribution corresponding to Eq.~(\ref{eq:superstat_5}). Color codes for varying $m$ and $n$ parameters. \leg{(b):} Wind speed distribution from the SIRTA observatory measurements (\textcolor{red}{$\bigcirc$}, $u$, and \textcolor{blue}{$\times$}, $v$) and its fit by the superstatistical distribution given by Eq.~(\ref{eq:superstat_5}) with parameters $m=9$, $n=0.019$ (magenta dashed line).}
\label{fig:SIRTA_superstat_module}
\end{center}
\end{figure}

We now turn to wind speed distribution, which is at stake for wind energy. For the sake of simplicity, we assume that the components $u$ and $v$  are independent statistically. Besides, we had previously observed that: (i) there is no correlation between $u$ and $v$ (Fig.~\ref{fig:wind_pdf_components}a-inset); (ii) $u$ and $v$ have similar statistics, \emph{i.e.} both components are described by the same $m$ and $n$. Therefore the wind speed distribution is obtained by computing the joint distribution in radial coordinates and integrated over the angle. Therefore, we obtain
\begin{equation}
 p(M)= \frac{n\,M}{2\,B(\frac{1}{2},m-1)^2} B(\frac{1}{2},\frac{1}{2})F(m,\frac{1}{2},1,-\frac{n^2 M^2}{1+nM^4}),
\label{eq:superstat_5}
\end{equation}
where $B$ is the Euler integral of the first kind and $F$ is the hypergeometric function. Again, this yields a heavy tail distribution (Fig.~\ref{fig:SIRTA_superstat_module}a), but for the set of parameters that fits the data the most (Fig.~\ref{fig:SIRTA_superstat_module}b), the wide tails occur for very low probability, and Eq.~(\ref{eq:superstat_5}) fits very well the data.

\subsection{Physical origin of super-statistics: stationary solutions of stochastic differential equations}
For a more in-depth analysis of the physical processes, several studies model the wind speed or its components using stochastic differential equation (SDE) (e.g. Nfaoui et al. 2004, Bernardin et al. 2009). A first order SDE for stochastic process $u$ can be written as:
\begin{equation}
\frac{d u}{d t} = \zeta\left( u \right) + \xi\left( u \right) \eta(t) \\
\label{eq:SDEx}
\end{equation}
where $\zeta$ and $\xi$ are arbitrary functions and $\eta$ is a random function of time, often referred to as a "noise term". Considering the time evolution of the probability density function, $p$, associated to the stochastic variable, $u$, Eq.~(\ref{eq:SDEx}) can be converted into the Fokker-Planck equation,
\begin{equation}
\frac{\partial p}{\partial t} = -\frac{\partial}{\partial u} \left( \zeta p \right) +
\frac{1}{2} \frac{\partial}{\partial u} \left( \xi \frac{\partial}{\partial u}
\left(\xi p \right) \right)  .
\label{eq:Fokker-Langevin_x}
\end{equation}
If $\xi$ is constant, the system is said to be subject to additive noise, otherwise it is said to be subject to multiplicative noise. The stochastic processes $u$ is usually a Markov process, i.e. characterized as memoryless: the next state depends only on the current state and not on the sequence of events that preceded it. A number of authors have modeled explicitly the temporal dependence of the wind speed through Markov chain models  (Youcef Ettoumi et al. 2003, Nfaoui et al. 2004, Shamshad et al. 2005, Fawcett and Walshaw 2006a, Fawcett and Walshaw 2006b). Interestingly, it turns out that the distribution of Eq.~(\ref{eq:superstatistics_4}) is also the stationary solution of Eq.~(\ref{eq:Fokker-Langevin_x}), with $\zeta(u)= -\gamma u$ and $\xi(u)= \sqrt{2 \gamma_M} |u| + \sqrt{2\gamma_A}$,
where $\gamma_M$  and $\gamma_A$ are constants governing the strength of the multiplicative (first term) and additive (second term) noise. The relations between $\gamma_M$, $\gamma_A$ and the parameters of Eq.~(\ref{eq:superstatistics_3}) are:
\begin{eqnarray}
\begin{array}{cc}
\displaystyle m & = \displaystyle \frac{\gamma+\gamma_M}{2\gamma_M}
\\
\displaystyle n & = \displaystyle \frac{\gamma_M}{\gamma_A}
\end{array}
\label{eq:relation_superstatistics_3_SDEx_2}
\end{eqnarray}
The multiplicative and additive noise terms can be physically interpreted as an interplay between turbulence, chaotic atmospheric variability and the mean wind (Bernardin et al. 2009). In the case of non-multiplicative noise, $\gamma_M=0$, the strength of the noise does not depend on the mean wind and the resulting stationary distribution is Gaussian. Conversely, the  multiplicative stochastic process, used to model wind stress over the ocean by Sura and Gille (2003), makes the link between the wind variability and its mean state and induces a stationary power-law distribution (Sakaguchi 2001). Yet, as the observed wind component distributions are narrower than power-laws, this multiplicative stochastic model is crude to describe them. Finally, besides the rather accurate but empirical fit of the Weibull distribution to measured wind speed distribution, is there thus a physical reason to model wind speed statistics with a Weibull distribution?
In order to address this question, a first step might be to consider a SDE with $\zeta(u)=-\gamma u$ and $\xi(u)= \sqrt{2 \gamma_M} |u|^k$ (Queir\'os 2008). Indeed, this yields the joint PDF for wind speed, $M$ and direction, $\phi$,
\begin{eqnarray}
 \begin{array}{ll}
 \displaystyle p(M,\phi) & = \displaystyle \frac{1}{Z^2} \left( \sin^2 \phi
 \cos^2 \phi \right)^{-k} M^{(1-2k)} \\
 \displaystyle & \displaystyle \times \exp\left[ -\frac{\gamma}{2\gamma_M
 (1-k)}M^{(2-2k)}  \left(\sin^{2(1-k)} \phi + \cos^{2(1-k)} \phi \right)
 \right].
 \end{array}
\label{eq:superstatistics_pdf_polar}
\end{eqnarray}
with $Z$ the corresponding normalization factor. Despite the fact that Eq.~(\ref{eq:superstatistics_pdf_polar}) can not be integrated with respect to $\phi$ between 0 and $2\pi$ because of integration singularities, it shares common features with a Weibull distribution, namely the power-law and stretched-exponential with exponent $k' = 2(1-k)$. Future works will thus explore in more detail the potential of such SDE to model wind speed statistics.

\section{Conclusion \label{sec_concl}}
The use of the Weibull distribution for wind statistics modeling and wind energy evaluation is a convenient and powerful approach. Is is however based on empirical more than physical justification and might display at times strong limitations for its application. In this article, we model the wind components instead of the wind speed itself, and thereby provide more physical insights on the validity domain of the Weibull distribution as a possible relevant model for wind statistics.

Based on wind measurements collected nearby Paris in a rather flat environment and in Southern France in the Rh\^one valley in a ridge and mountains environment, this article proposes different probability density functions of the wind components. This opens the path to the analysis of the applicability of the Weibull distribution as a suited model for wind statistics and the quantification of the errors on the mean wind speed and wind power arising from the use of the Weibull distribution.

This study first shows that in complex terrain, the existence of terrain induced wind system affects the surface wind distribution which does not display a Weibull shape. Modeling the wind statistics with a Weibull  distribution produces errors exceeding $10\%$ on the wind power estimate. The combination of a Rayleigh distribution to model the ``isotropic'' wind and a Rice distribution to model persistent wind regimes gives excellent results for the three surface weather stations located in the Rh\^one valley. Combining Rayleigh and Rice distributions is equivalent to applying the recently developed super-statistics theory, which models the wind system as the superposition of different local dynamics at different intervals with different mean wind speeds. However, Rayleigh and Rice distributions assume normal distributions for the two wind components. The distributions of the wind components, even over flat terrain, however exhibit non Gaussian distribution. This can also be modeled using super-statistics theory, which assumes fluctuating variances of the two wind components, that are eventually modeled by a Gaussian distribution over a short time interval. However, this produces power-law distribution for large wind speed value which contradict observed distributions and Weibull model. This can be partly overcome by assuming the wind components to verify stochastic differential equation with multiplicative noise with power coefficient.

Parametric statistical downscaling using the Weibull distribution is often used to produce regional surface wind speed climatologies (Pryor et al. 2005). Also, this study points out the limits of the use of parametric distribution to downscale surface wind speed and advocates for non parametric statistical models (e.g. Michelangeli et al. 2009, Salameh et al. 2009, Lavaysse et al. 2012,Vrac et al. 2012).

\ack
The authors are very grateful to Samuel Humeau and Jerry Szustakowski for fruitful discussions.

\end{document}